\documentclass[twocolumn, 
prb,preprintnumbers,amsmath,amssymb,floatfix]{revtex4}

 \usepackage[linktocpage,bookmarksopen,bookmarksnumbered]{hyperref}
\usepackage{graphicx}
\usepackage{dcolumn}

\begin{document}

\title{Crystal Symmetry and Magnetic Order in Iron Pnictides: \\
a Tight Binding Wannier Function Analysis}
\author{Z. P. Yin}
\thanks{Present address: Department of Physics and Astronomy, Rutgers University, Piscataway, NJ 08854.}
\author{W. E. Pickett}
\affiliation{Department of Physics, University of California Davis, 
  Davis, CA 95616}
\date{\today}
\begin{abstract}
To perform a local orbital analysis of electronic and magnetic interactions,
we construct the  Wannier functions (WFs) of the Fe $3d$ orbitals in the parent compound of the 
recently discovered iron pnictide superconductors, LaFeAsO, and a comparison material
LaFePO.  Comparing the WFs for the stripe antiferromagnetic 
order with those for no magnetic order, the difference is a significant spreading (``{\it de}localization'')
of specifically the $d_{xy}$ and $d_{xz}$ (but not $d_{yz}$) WFs, where parallel Fe spins lie 
along the $x$ direction.
The WF basis gives a tight-binding representation of the first principles, density functional based 
Fe-derived bands. Comparing hopping parameters, it is found that
changes due to stripe antiferromagnetism, even if it is weak, enables more isotropic hopping involving
spin-majority electrons in the Fe $3d_{xz}$ (but not the $3d_{yz}$) orbital.
This change, counterintuitively, actually reinforces electronic anisotropy.
Further insight is gained by comparing the WFs of LaFeAsO and LaFePO, identifying how the 
difference in WFs is related to the difference in hopping integrals and
showing how the pnictide atom is influential in forming the stripe antiferromagnetism.
Kinetic energy considerations suggest that orbital fluctuation, in addition to spin fluctuation,
may contribute to the decrease in observed ordered moment compared to the calculated values.   
\end{abstract}
\maketitle

\section[Background and Motivation]{Background and Motivation}
Since the first report from Hosono's group\cite{1stLaFeAsO} of superconductivity at 
$T_c$=26 K in  F-doped LaFeAsO,
hundreds of experimental and theoretical papers on these iron-pnictide compounds have appeared, 
aimed at elucidating various properties, including synthesizing new compounds to achieve higher T$_c$, 
measuring basic quantities (e.g. magnetic susceptibility, NMR, ARPES), 
and modeling and simulating to obtain explanations and predictions.
Thanks to these efforts, there are now several families of these iron pnictide superconductors, 
including the 1111-family (e.g. LaFeAsO, CaFeAsF), 122 family (e.g. BaFe$_2$As$_2$), 
111-family (e.g. LiFeAs) and a more complicated 22426-family (e.g. Fe$_2$As$_2$Sr$_4$Sc$_2$O$_6$),
with T$_c$ up to 56 K.[\onlinecite{Wang-GdO}] 
Several aspects have been clarified: the superconductivity lies in primarily iron $3d$ bands\cite{Liu-isotope} 
and is not phonon-mediated;\cite{boeri}
the ground state in most classes 
is a stripe antiferromagnetic phase 
with a significantly reduced Fe magnetic moment compared to theoretically calculated value;\cite{NIST,Yin:arXiv0804.3355} 
it is a moderately correlated system where a Coulomb interaction U$\approx$3 eV might be
appropriate.[\onlinecite{Biermann}]
There is discussion that the superconducting order parameter may have a new $s_{\pm}$ character.\cite{mazin, Nagai}

Despite a great deal of progress in understanding the electronic structure\cite{SinghDu,Yin-njp,MazJoh}
and magnetic interactions,\cite{Yaresko,Han,MDJ} some basic questions remain unresolved.
One of them is:
what is the underlying mechanism of the structural transition from tetragonal to orthorhombic 
in the parent compounds of iron-based superconductors?
This question is especially challenging in the 1111-compounds (e.g. LaFeAsO), where the structural transition
is observed (as the temperature is lowered) 
to occur before\cite{mcguire} the magnetic transition (from nonmagnetic to stripe 
antiferromagnetic order which we denote as Q$_M$ AFM).
It would have been natural to think that the stripe antiferromagnetic ordering of Fe provides the driving force for
the structural transition because it introduces electronic anisotropy. 
(Table III in reference [\onlinecite{Ishida-summary}] provides a summary of the structural transition temperature T$_S$ 
and stripe antiferromagnetic transition temperature T$_N$ of several iron pnictide compounds.)

Noting that the structural transition and magnetic transition occurs simultaneously in
the 122 compounds (e.g. BaFe$_2$As$_2$),
a possible argument is that the magnetism is in fact present, in the form of medium-range
order, antiphase boundaries, etc., near the structural transition but its detection is greatly suppressed
by strong spatial or temporal fluctuation.  The suggestion by Mazin and Johannes that magnetic antiphase boundaries
may be the dominant excitation\cite{iimmdj} has already stimulated numerical estimations by the present 
authors.\cite{zpywep}
With a time resolution of $10^{-15}$ s, 
photoemission experiments by Bondino {\it et al.}\cite{Bondino} implied a dynamic magnetic moment 
of Fe with magnitude of 1 $\mu_B$ 
in the nonmagnetic phase of CeFeAsO$_{0.89}$F$_{0.11}$, 
which is comparable to the ordered magnetic moment of Fe in the undoped antiferromagnetic CeFeAsO compound. 
The fluctuation strength should be much stronger in 1111-compounds than 122 compounds 
based on the fact that the measured
Fe ordered magnetic moment in 1111-compounds ($\sim$ 0.4 $\mu_B$) is much less than in 122 compounds ($\sim$ 0.9 $\mu_B$)
and they are much smaller than DFT predicted value ($\sim$ 2 $\mu_B$).\cite{Ishida-summary, Yin:arXiv0804.3355} 
One factor is that interlayer coupling of FeAs layers is stronger in 122 compounds than 1111-compounds
because the interlayer distance in 122 compounds ($\sim$6 $\AA$) 
is significantly smaller than 1111-compounds ($\sim$8-9 $\AA$).\cite{Ishida-summary} 
Interlayer interaction should help to stabilize the ordered Fe magnetic moment 
by reducing fluctuations (reducing two dimensionality).

In this paper we address the effect of magnetic order, and of the pnictide atom, on the 
strength, character, and spin-dependence of
Fe-Fe hopping processes by using a Wannier function representation based on all five Fe $3d$ orbitals,
and only these orbitals.  Several previous studies of the electronic structure have pointed out
some aspects of the influence of the pnictide, or chalcogenide, atoms (due to size or chemical 
identities) and also of their positions.\cite{Yin:arXiv0804.3355,
Vildosola,Belash,Yildirim, Berciu}  We 
provide one example of the effect of the pnictogen atom (comparing LaFaAsO with LaFePO) in this paper,
where the effect of the pnictogen is included precisely but indirectly through the Wannierization process.
This allows us to present results in 
an Fe-centric picture. This local orbital representation provides insight
into both electronic and magnetic behavior even when the fundamental behavior is primarily itinerant.  

\begin{figure}[bht]
\vskip 5mm
\rotatebox{-90}
{\resizebox{6.0cm}{8.0cm}{\includegraphics{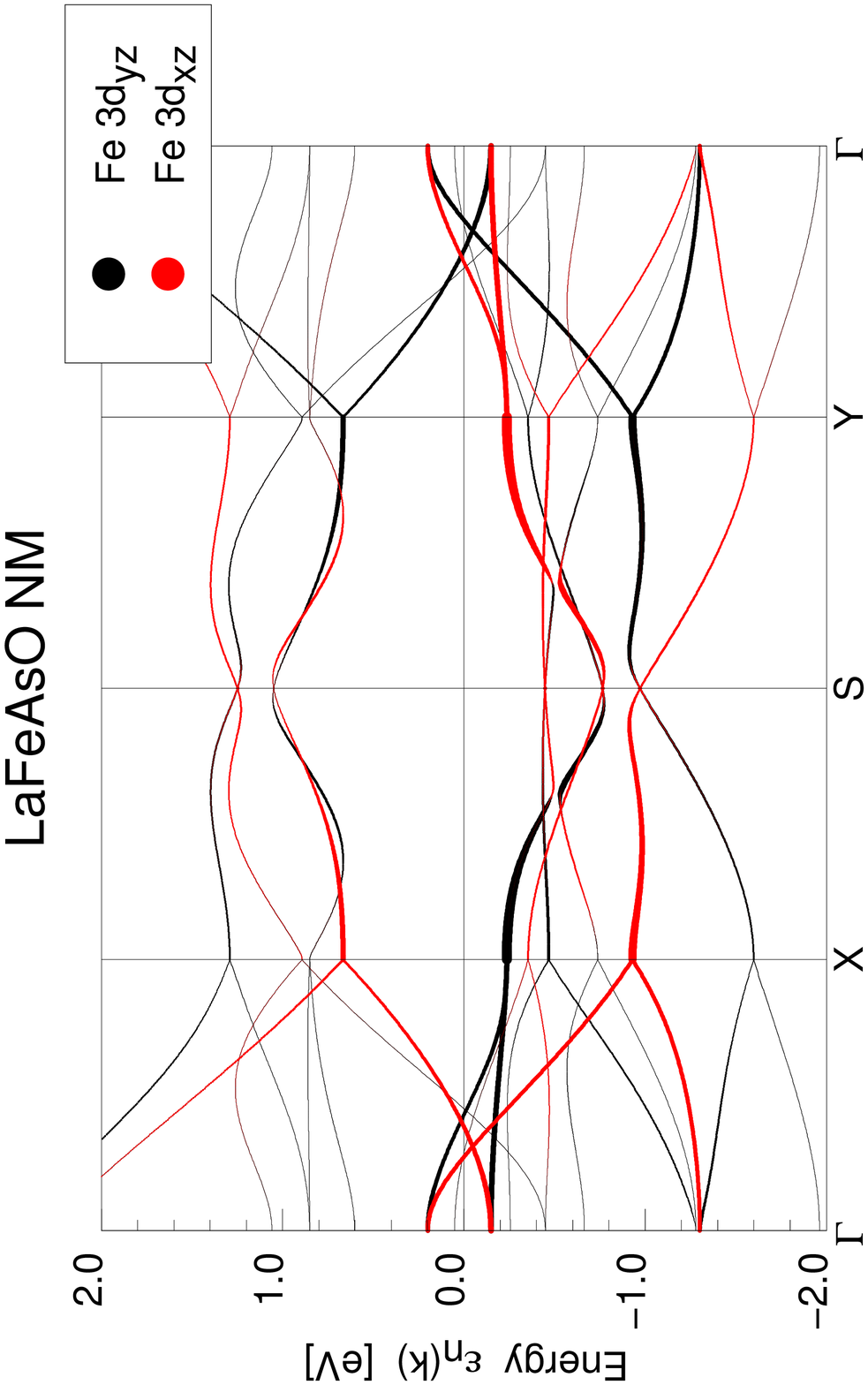}}}
\vskip 6mm
\rotatebox{-90}
{\resizebox{6.0cm}{8.0cm}{\includegraphics{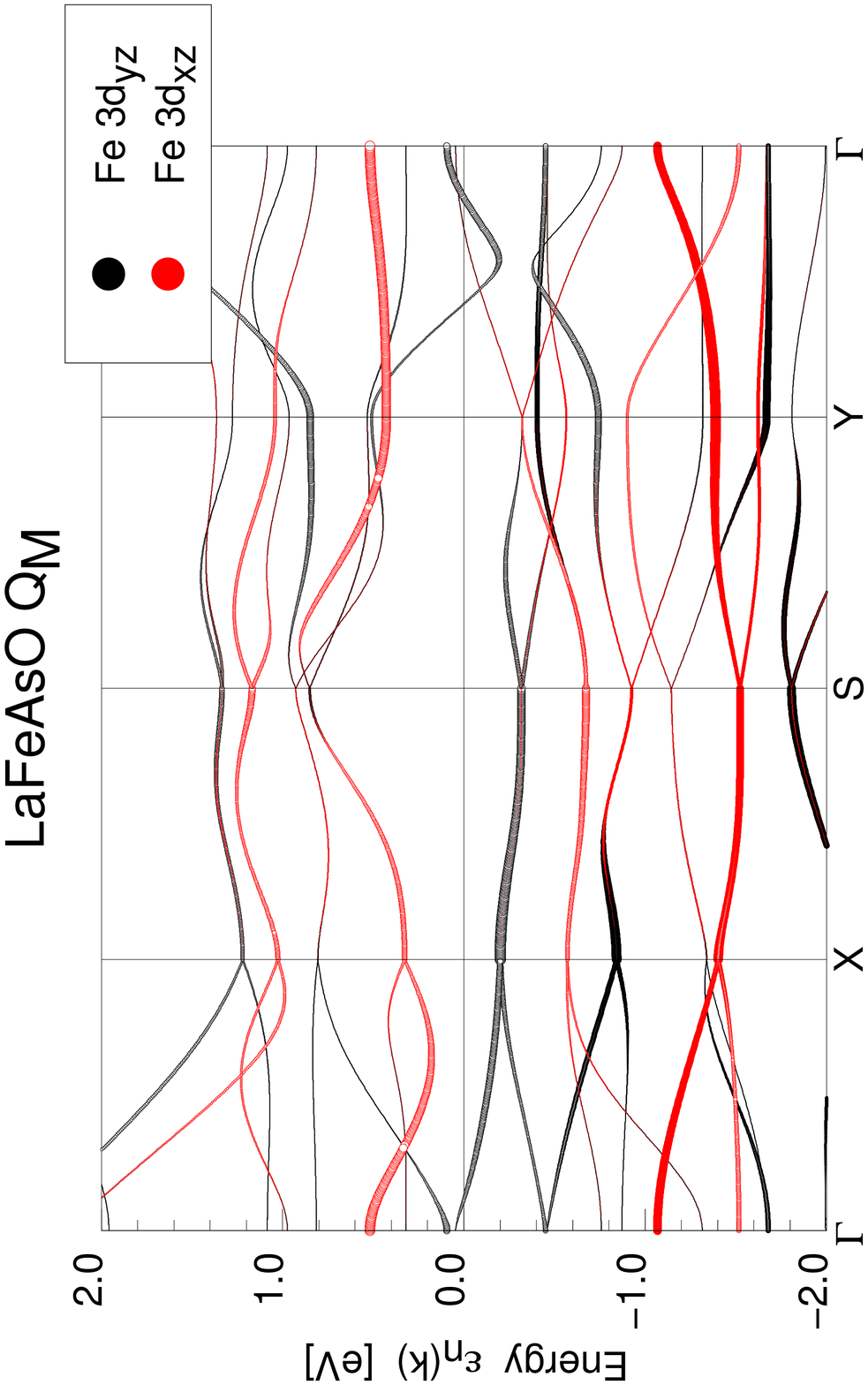}}}
\caption{LaFeAsO band structure with highlighted Fe $3d_{yz}$ and $3d_{xz}$ fatband characters 
in the NM (top panel) and Q$_M$ AFM (bottom panel) phases. Compared to the NM phase, 
the Fe $3d_{xz}$ bands near Fermi level in the Q$_M$ AFM phase,
especially along $\Gamma -X$ and $\Gamma -Y$ directions,
change dramatically due to the formation of the stripe antiferromagnetism with large ordered Fe magnetic moment of
1.9 $\mu_B$. }
\label{As-fatband}
\end{figure}

\begin{figure}[htb]
\rotatebox{-90}
{\resizebox{6.0cm}{8.0cm}{\includegraphics{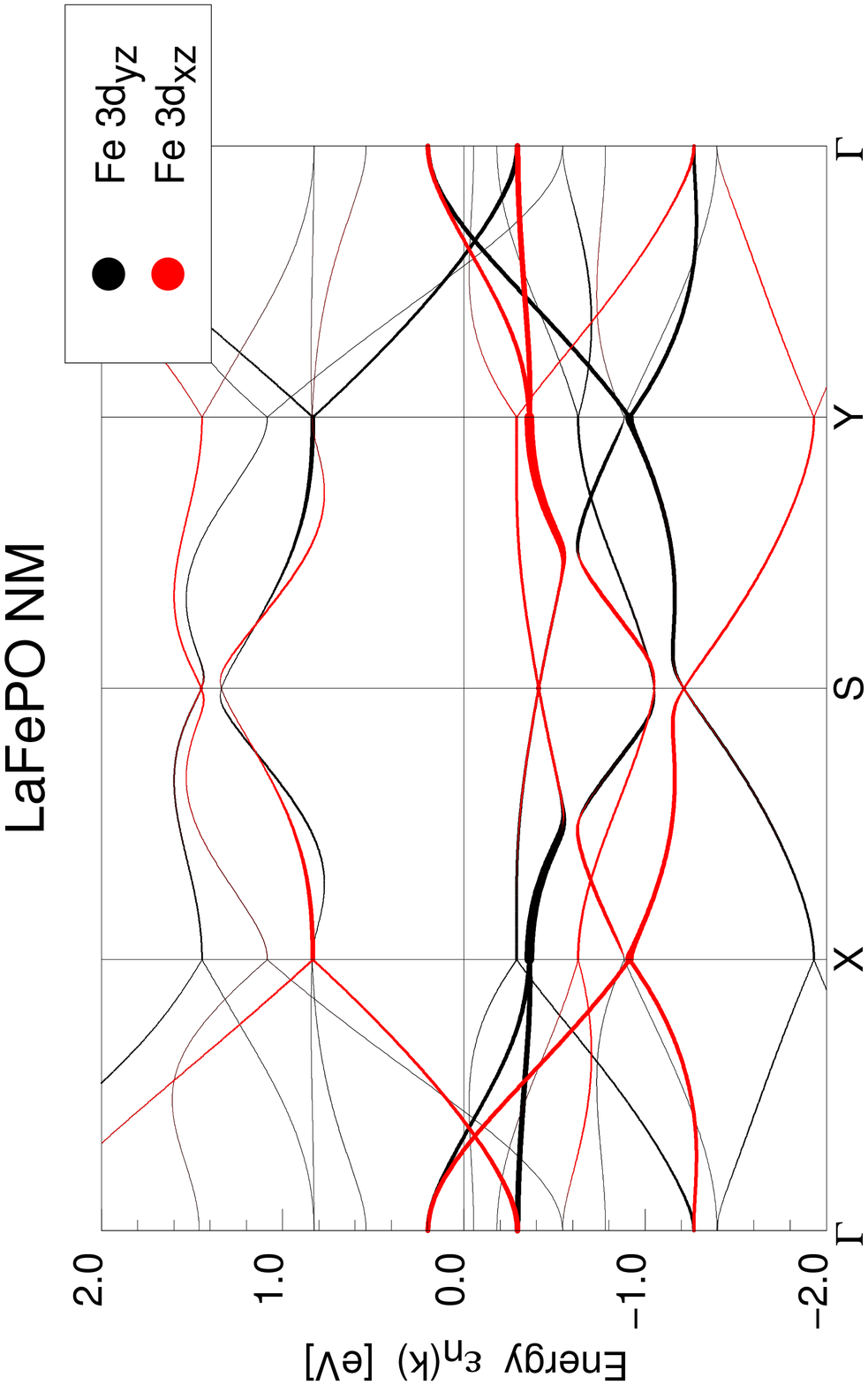}}}
\vskip 6mm
\rotatebox{-90}
{\resizebox{6.0cm}{8.0cm}{\includegraphics{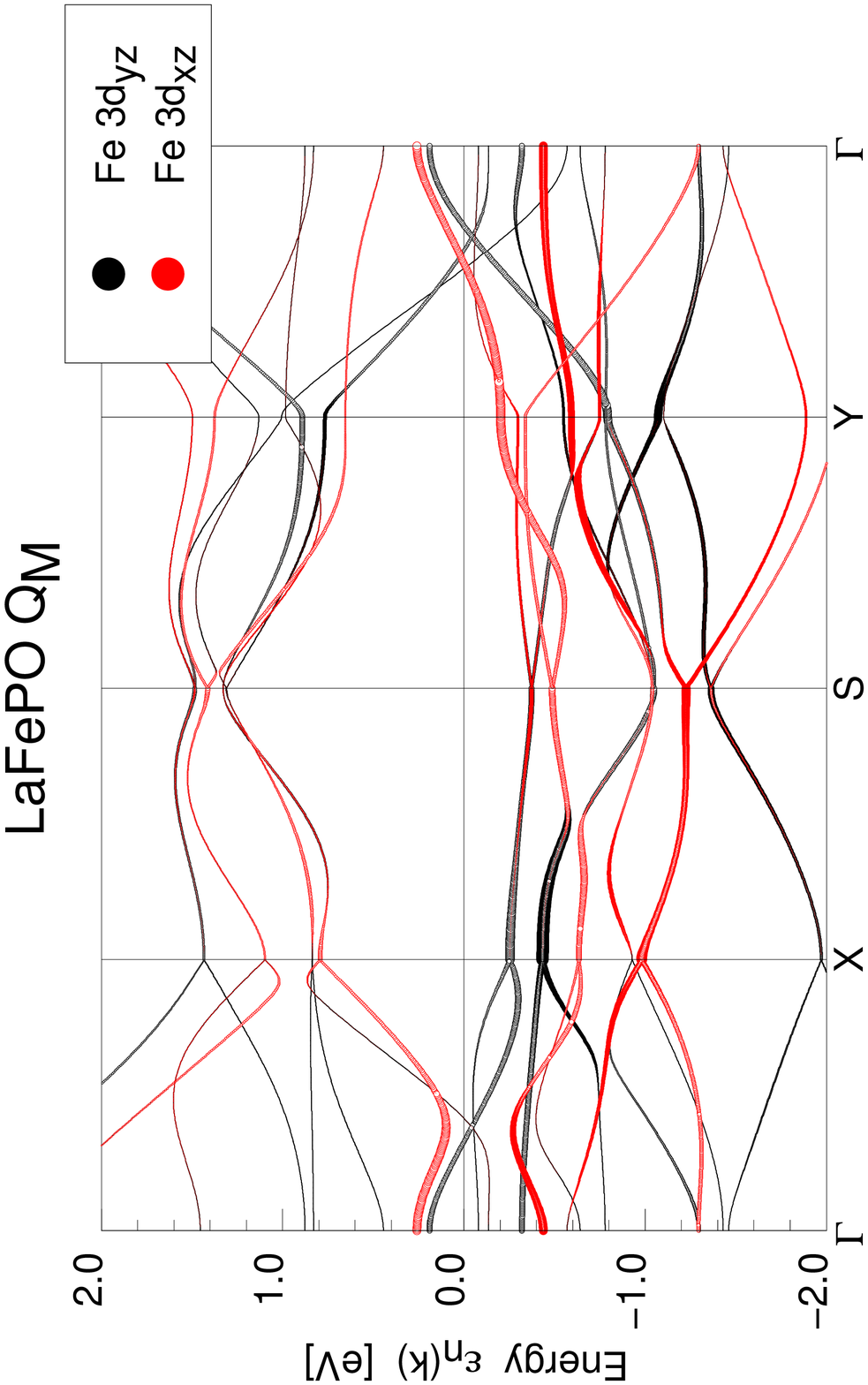}}}
\caption{LaFePO band structure with highlighted Fe $3d_{yz}$ and $3d_{xz}$ fatband characters
 in the NM (top panel) and Q$_M$ AFM (bottom panel) phase. 
Compared to LaFeAsO, 
the Fe $3d_{xz}$ bands near Fermi level in the Q$_M$ AFM phase change less significantly from the NM phase,
 due to the relatively small ordered Fe magnetic moment of 0.5 $\mu_B$. }
\label{P-fatband}
\end{figure}

\section[Calculational methods]{Calculational methods}
We begin with  first principle calculations using the full-potential local-orbital code\cite{FPLO} (FPLO8)
with local density approximation (LDA) exchange-correlation (XC) functional\cite{PW91} (PW92) and 
the experimental lattice constants and internal atomic coordinates for the compounds
LaFeAsO and LaFePO, 
as given in our previous work\cite{Yin:arXiv0804.3355, Yin-njp, zpywep}.
To obtain a consistent local orbital representation and the resulting hopping amplitudes, 
we then construct real-space Wannier functions (WFs) derived from
Fe $3d$ orbitals in both NM and Q$_M$ AFM phases.
The WFs used in this paper, as implemented in the FPLO8 code, are constructed by projecting the Bloch functions
from a specified energy range onto chosen atomic orbitals, roughly following the method of
Ku {\it et al.}\cite{Ku1,Ku2}  The resulting Wannier orbitals retain a symmetry that is common to both the atomic
orbital and the point group symmetry of the site.
These WFs provide an explicit basis set of local orbitals that give a
tight binding representation, complete with on-site energies and hopping amplitudes to
neighbors as distant as necessary to represent the chosen bands.  In this paper we project onto the conventional
real Fe $3d$ orbitals, with the energy range corresponding to the region with strong Fe $3d$ character in the bands.

\section{Differences in Band Structures}
The differences in electronic structure that we will emphasize result from the changes due to stripe magnetic order, and
the differences between LaFeAsO with larger ordered moment, and LaFePO, with smaller calculated moment (experimentally
nonmagnetic).  The necessary band structures are shown in Fig. \ref{As-fatband} for LaFeAsO and \ref{P-fatband} for LaFePO,
where in each case the Fe $3d_{yz}$ and $3d_{xz}$ characters are highlighted.
The total energy of LaFePO, which is experimentally found to be nonmagnetic, is only slightly lower (2 meV/Fe) in the Q$_M$ 
AFM phase than the nonmagnetic phase,\cite{Yin-njp} so the incorrect prediction for LaFePO is actually a fine detail,
and suggests it is nearly antiferromagnetic.
The calculated Fe magnetic moment of LaFePO in the Q$_M$ AFM phase is 0.52 $\mu_B$.  In LaFeAsO, the calculated 
moment is near 1.9$\mu_B$, substantially larger than 
the measured value of 0.36 $\mu_B$ as has been widely discussed (see, for example, Refs. [\onlinecite{Yin-njp,NIST}]).

\begin{figure}[tbp]
{\resizebox{9.0cm}{10.0cm}{\includegraphics{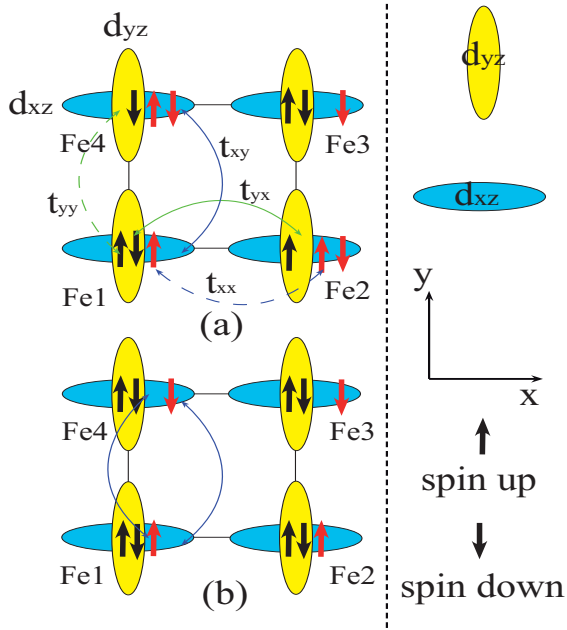}}}
\caption{(color online)Possible orbital orderings of iron in iron-pnictides. Left panel: Both (a) and (b) form
the Q$_M$ AFM ordering. However, (a) is favored because it gains more kinetic energy from nearest-neighbor hoppings
according to second-order perturbation theory (see text). 
Right panel (from top to bottom) shows the simplified symbols for Fe $3d_{yz}$ and
$3d_{xz}$ orbitals, the chosen $x$ and $y$ directions, up arrows for spin up electrons 
and down arrows for spin down electrons, where black arrows for $3d_{yz}$ orbital and red arrows for $3d_{xz}$ orbital. }
\label{hopping}
\end{figure}

For our calculations and discussion we have chosen the $x$-axis along the direction of aligned Fe spins, 
as shown in Fig. \ref{hopping}; the corresponding zone boundaries are denoted X and Y in the band plots. 
The nonmagnetic band structures of the two compounds are very similar, differing only in some fine details that 
do not arise in our analysis.  
However, the band structures in the Q$_M$ AFM phase of the two compounds differ substantially, 
which is due to the difference in the Fe magnetic moment (1.9 vs. 0.5 $\mu_B$).\cite{Yin:arXiv0804.3355, Yin-njp}
The similarities and differences provide a way to study the effect of magnetic order in these compounds, and 
specifically to show that even small magnetic order has substantial consequences.  Since the AFM and NM phases in
LaFePO are nearly degenerate, our results have relevance to the effect of (longitudinal) magnetic fluctuations of the Fe atom.

The panels in Fig. \ref{As-fatband} illustrate that the magnetic order substantially simplifies the band structure
very near the Fermi level, which is all near $\Gamma$ in this doubled (magnetic) cell.  The other difference to notice
is the great difference in band structure along $\Gamma$-X and $\Gamma$-Y directions.
Figure \ref{P-fatband} shows the influence of a weak stripe antiferromagnetism (0.5 $\mu_B$) on the nonmagnetic band structure.
The overall band structure remains the same except for some bands near the Fermi energy, 
where the main change is the separating of the Fe $3d_{xz}$ bands away from the Fermi level,
which causes disappearance and change of topology of certain pieces of the Fermi surface of the Fe $3d_{xz}$ bands.
Note that the Fe $3d_{yz}$ bands change insignificantly, leaving the bands near the
Fermi level dominated by Fe $3d_{yz}$ character. 
This difference indicates that even a weak stripe antiferromagnetism has a very strong symmetry breaking
effect on the $3d_{xz}$ and $3d_{yz}$ bands, which are equivalent in the nonmagnetic state.
As a result, even a weak stripe antiferromagnetism induces a large anisotropy, let alone the much stronger (calculated) 
antiferromagnetism in FeAs-based compounds. 
(The much bigger anisotropy in the stripe AFM phase in LaFeAsO is evident by comparing Fig. \ref{As-fatband} and \ref{P-fatband}.) 

\section{$3d_{xz}$ and $3d_{yz}$ Orbital Repopulation}
Due to the strong influence of stripe antiferromagnetism on the band structure (even when weak as in LaFePO), 
the orbital distinction and repopulation of the Fe $3d_{xz}$ and $3d_{yz}$ electrons suggests various
means of analysis.
The strong intra-atomic anisotropy discussed above is sometimes referred to as
orbital ordering, but with the orbital occupations far from integers, the anisotropy also has a substantial
itinerant (collective) component.  Here we consider briefly the alternative, local viewpoint.

Figure \ref{hopping} shows two underlying (idealized) orbital populations, 
both of which are consistent with the Q$_M$ AFM symmetry.
(This orbital differentiation is often called ``orbital ordering,'' but based on the calculated populations,
discussed below, this is more properly thought of as an itinerant cousin of orbital ordering.) 
$t_{xy}$ denotes the hopping parameter of the $d_{xz}-d_{xz}$ hopping in the $y$ direction, and $t_{yx}$ the
$d_{yz}-d_{yz}$ hopping in the $x$ direction. In the nonmagnetic case, by symmetry
\begin{equation}
t_{xy}=t_{yx}=t,
\end{equation} 
whereas they differ in the Q$_M$ AFM state. 
$t_{xx}$ will denote 
$d_{xz}-d_{xz}$ hopping 
in the $x$ direction, and similarly $t_{yy}$ denotes $d_{yz}-d_{yz}$ hopping in the $y$ direction (see Fig. \ref{hopping}).

Let $U$ and $U'$ denote the intra-orbital and inter-orbital Coulomb repulsion, and 
$J_H$ the inter-orbital Hund's exchange constants. Our purpose is to estimate the 
difference in kinetic energy gain of the two configurations shown in Fig. \ref{hopping}.
At the level of second-order perturbation theory, the kinetic energy gain from the 
$d_{yz}-d_{yz}$ hopping in the $x$ direction  
(Fig. \ref{hopping}a) is 
\begin{equation}
\Delta E_{yx}=-t_{yx}^2/(U'-J_H).
\end{equation} 
A similar kinetic gain of 
\begin{equation}
\Delta E_{xy}=-t_{xy}^2/(U'-J_H)
\end{equation}
comes from the $d_{xz}-d_{xz}$ hopping in the $y$ direction 
(Fig. \ref{hopping}a). 
$t_{xx}$ and $t_{yy}$ are much smaller and can be neglected (see Table \ref{LaFeAsO-hopping}).
Therefore, the total energy
gain from NN hopping of Fig. \ref{hopping}a is 
\begin{equation}
\Delta E(a) = \Delta E_{xy} +\Delta E_{yx} =-2t^2/(U'-J_H),
\end{equation}
while it is 
\begin{equation}
\Delta E(b) =-2t^2/U
\end{equation} for Fig. \ref{hopping}b.
Because $U$ is larger than $U'-J_H$, the orbital ordering in Fig. \ref{hopping}(a) is 
favored over Fig. \ref{hopping}(b) 
by kinetic fluctuations. 
This result is a more transparent form of an analysis presented by 
Lee {\it et al.}\cite{Lee}. 

\begin{figure}[tbp]
{\resizebox{8cm}{6cm}{\includegraphics{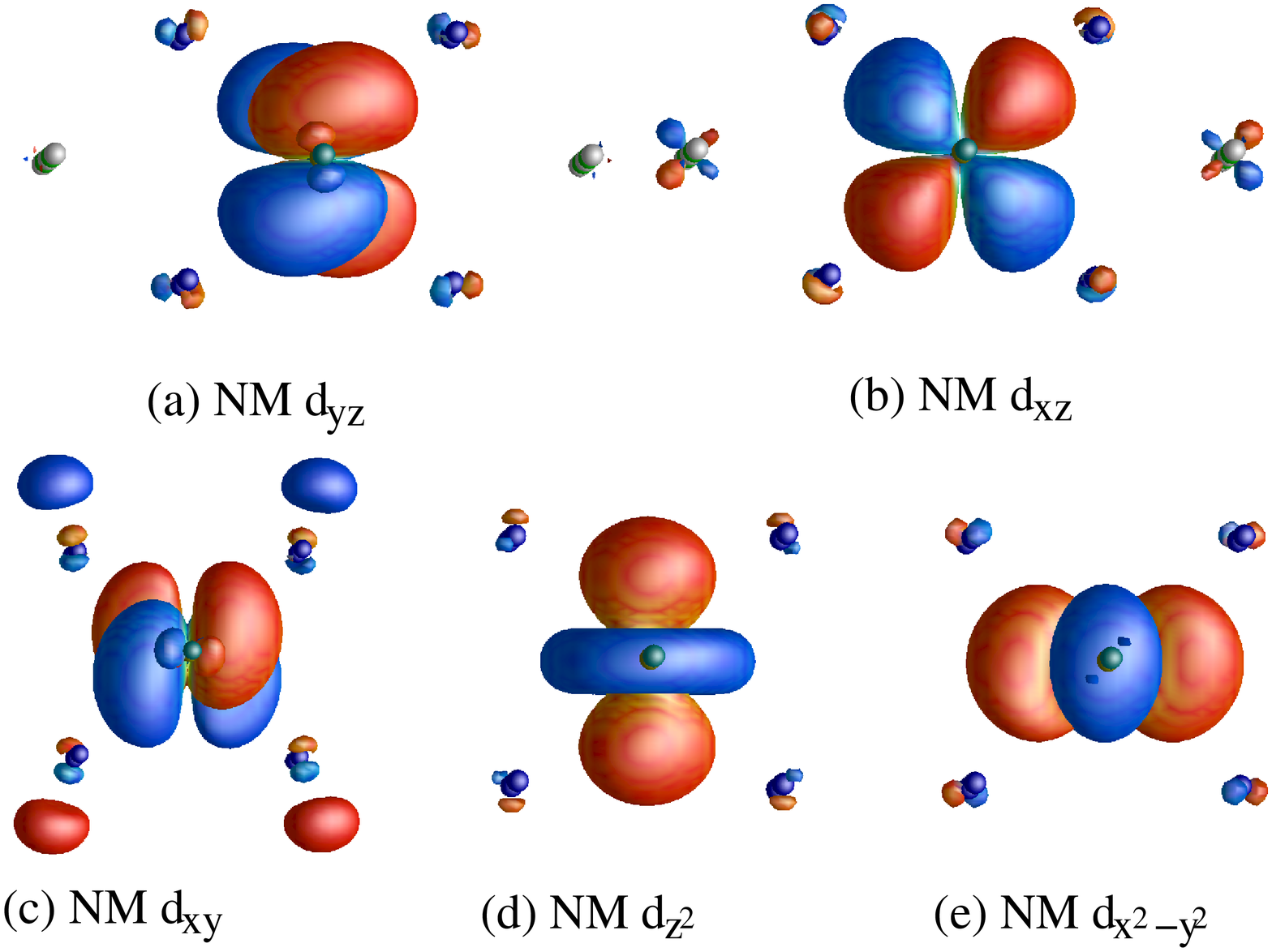}}}
\vskip 5mm
{\resizebox{8cm}{6cm}{\includegraphics{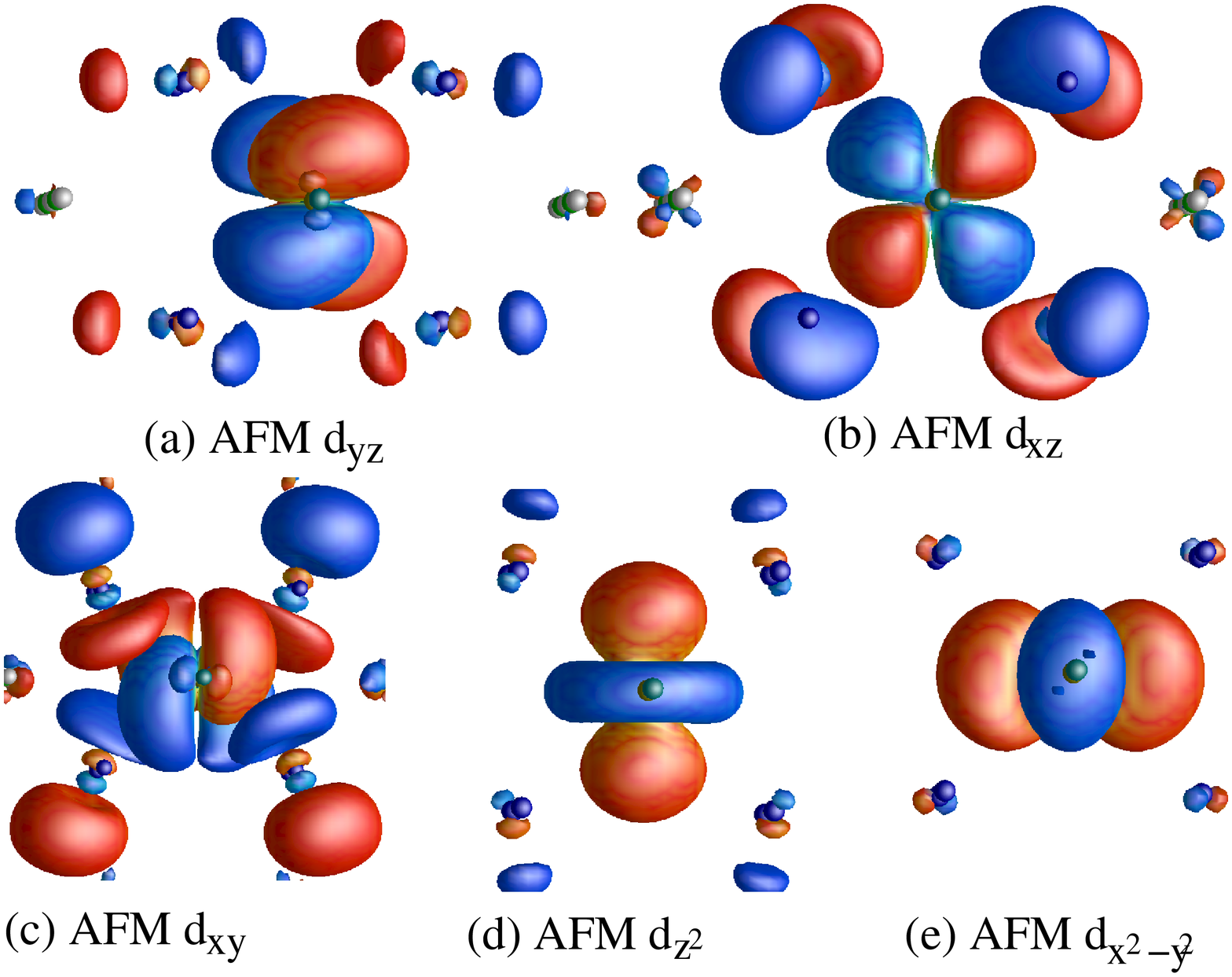}}}
\caption{LaFeAsO Wannier functions of Fe $3d$ orbitals in the NM phase (top panel), and those 
for the majority spin in Q$_M$ AFM phase (bottom panel).
The Wannier functions of Fe $3d$ orbitals for the minority spin in the Q$_M$ AFM phase 
remain almost the same as in the NM phase.
The important difference to be observed is that
in the Q$_M$ AFM phase, the (majority spin) Wannier functions of $3d_{xz}$ and $3d_{xy}$ orbitals 
(and only these) are more extended, with much increased density at neighboring As sites.
The isosurface has the same value (density) in each panel.
 }
\label{LaFeAsO-wan-NM-QM}
\end{figure}

\begin{table*}[htb]
\caption{The hopping parameters (in eV) 
of the Fe1 $3d_{yz}$, $3d_{xz}$, and $3d_{xz}$ orbitals
to all the five $3d$ orbitals of its nearest neighbor Fe2 and Fe4 atoms and next-nearest neighbor Fe3 atom 
in the nonmagnetic and Q$_M$ AFM phases of LaFeAsO. The highlighted (italicized and boldface) entries are 
discussed in the text.
 }
\label{LaFeAsO-hopping}
\begin{tabular}{|cc|ccc|ccc|ccc|}
\hline
Fe1   &  & \multicolumn{3}{|c|}{$yz$} & \multicolumn{3}{|c|}{$xz$} & \multicolumn{3}{|c|}{$xy$} \\
\hline
      &  & NM & \multicolumn{2}{c|}{Q$_M$} & NM & \multicolumn{2}{c|}{Q$_M$} & NM & \multicolumn{2}{c|}{Q$_M$} \\
      &  &    & up    & dn    &    & up    & dn    &    & up    & dn \\
\hline
Fe2  & $z^2$     & -0.12 & -0.16 & -0.08 &  0    &  0    &  0   & 0 & 0 & 0 \\
     & $x^2-y^2$ &  0.34 &  0.42 &  0.28 &  0    &  0    &  0   & 0 & 0 & 0  \\
     & $yz$      & {\it -0.33} & {\it -0.42} & {\it -0.29} &  0    &  0    &  0 & 0 & 0 & 0    \\
     & $xz$      &  0    &  0    &  0    & -0.06 & {\bf -0.29} &  0.09  & -0.22 & -0.21 & -0.20 \\
     & $xy$      &  0    &  0    &  0    & -0.22 & -0.21 & -0.20  & {\bf -0.18} & {\bf -0.33} & {\bf -0.07} \\
\hline
Fe4  & $z^2$     &  0    &  0    &  0    & -0.12 & -0.11 & -0.15 & 0 & 0 & 0 \\
     & $x^2-y^2$ &  0    &  0    &  0    & -0.34 & -0.39 & -0.34 & 0 & 0 & 0  \\
     & $yz$      & -0.06 & -0.09 & -0.09 &  0    &  0    &  0    & -0.22 & -0.21 & -0.20 \\
     & $xz$      &  0    &  0    &  0    & {\it -0.33} & {\it -0.35} & {\it -0.35} & 0 & 0 & 0 \\
     & $xy$      & -0.22 & -0.20 & -0.27 &  0    &  0    &  0   & -0.18 & -0.23 & -0.23  \\
\hline
Fe3  & $z^2$     & -0.10 & -0.10 & -0.11 & -0.10 & -0.12 & -0.10  & -0.17 & -0.20 & -0.21 \\
     & $x^2-y^2$ &  0.10 &  0.09 & -0.10 & -0.10 & -0.09 & -0.09  & 0  & 0.02 & -0.02 \\
     & $yz$      & {\it 0.22} &  {\it 0.23} &  {\it 0.24} &  0.08 &  0.12 &  0.08  & -0.01 & 0.01 & 0 \\
     & $xz$      &  0.08 &  0.08 &  0.12 & {\it 0.22} & {\it 0.24} & {\it 0.24} & -0.01 & -0.02 & 0.03 \\
     & $xy$      &  0.01 &  0    & -0.01 &  0.01 & -0.03 &  0.02  & 0.13 & 0.13 & 0.13 \\
\hline
\end{tabular}
\end{table*}

\section[Tight Binding Hopping Parameters and Wannier Functions]{Tight Binding Hopping Parameters and Wannier Functions}

Figure \ref{LaFeAsO-wan-NM-QM} shows the Wannier functions (WFs) of all five Fe $3d$ orbitals in both
NM and Q$_M$ AFM (majority spin) phases of LaFeAsO, using the same value of isosurface in all cases. 
In the NM (spin-degenerate, tetragonal) phase, all five WFs for Fe $3d$ orbitals 
have their density strongly
concentrated on the Fe site.
All Fe minority spin $3d$ WFs in the Q$_M$ AFM phase remain 
almost the same as in the NM phase, 
so they are not shown.
The majority spin WFs for $3d_{yz}$, $3d_{x^2-y^2}$ and $3d_{z^2}$ orbitals remain
very similar to the corresponding Wannier functions in the NM phase, as can be seen
in Fig. \ref{LaFeAsO-wan-NM-QM}. 
The significant change is that the majority spin WFs for $3d_{xz}$ and $3d_{xy}$ become 
more delocalized in the Q$_M$ AFM phase,
with significant density at the nearest neighbor As sites, the effect being especially large 
for the $d_{xz}$ orbital.
This difference reveals that the majority spin Fe $3d_{xz}$ and $3d_{xy}$ orbitals  
mix much more strongly with nearest-neighbor As $4p$ orbitals in the Q$_M$ phase than in the NM phase.
The AFM order involves a highly anisotropic magnetization, and resulting 
difference in majority and minority potentials,
that produces this strongly orbital-dependent effect.

Using these WFs as the basis gives a tight binding 
representation, for which the hopping parameters are obtained from 
matrix elements of the Wannier Hamiltonian using the FPLO8 code. 
The corresponding band structures of LaFeAsO and LaFePO are already shown in 
Fig. \ref{As-fatband} and Fig. \ref{P-fatband} and the
resulting tight binding bands (not shown) fit very well the corresponding DFT-LSDA 
Fe-derived bands in both NM and stripe AFM phases.

Table \ref{LaFeAsO-hopping} presents the hopping parameters of the Fe1 $3d_{yz}$, 
$3d_{xz}$, and $3d_{xy}$ orbitals
to all the $3d$ orbitals of its nearest neighbor Fe2 and Fe4 atoms and next nearest 
neighbor Fe3 atom in LaFeAsO. 
(See Fig. \ref{hopping} for the definition of each Fe atom.)

The on-site energies (in eV) of all the five $3d$ orbitals in the NM phase and Q$_M$ AFM phase 
in both LaFeAsO and LaFePO are shown in Table \ref{onsite-energy}.
In the Q$_M$ AFM phase, the on-site energies are shown separately for both spin up 
(majority spin) and spin down (minority spin) orbitals.  For LaFeAsO, the $d_{xz}, d_{yz}$ energies lie at the
Fermi level, $d_{z^2}$ and $d_{x^2-y^2}$ lie 0.1-0.3 eV below while $d_{xy}$ lies about 0.2 eV
above, without magnetic order.  With AFM order, the minority on-site energies do not change
greatly, whereas the majority levels fall by 0.7-0.8 eV, thereby affecting the hybridization with
the As $4p$ orbitals.  The changes in LaFePO are smaller, corresponding to the smaller
(factor of $\sim$3) magnetic moments.

The hopping parameters reported here are similar to the corresponding hopping parameters 
reported by Lee {\it et al.}\cite{Lee} and Haule {\it et al.}\cite{Haule-hopping}.
(The differences between our results and those of Lee {\it et al.} reflect the fact that, although the
original bands are the same and the Wannier transformation is formally the same, the Wannier transformation 
is not unique and depends somewhat on some details in the implementation.)  
Our hopping amplitudes
are not directly comparable to those reported by Cao {\it et al.}\cite{hirschfeld} 
who focused on the hopping parameters from As $4p$ orbitals to Fe $3d$ orbitals and to its 
nearest neighbor As $4p$ orbitals.
As shown in Table \ref{LaFeAsO-hopping}, in the NM phase, $t_{xy}=t_{yx} >> t_{xx} =t_{yy}$, which indicates that 
the hopping (through As atoms) of $d_{xz}-d_{xz}$ ($d_{yz}-d_{yz}$) in the $y$ ($x$) direction of the electrons in
Fe $3d_{xz}$ ($3d_{yz}$) orbital is favored over the $x$ ($y$) direction. 
The hopping process for Fe $3d_{xz}$ ($3d_{yz}$) electrons is anisotropic.
Global tetragonal symmetry is retained because the Fe $3d_{xz}$ and $3d_{yz}$ electrons hop in different directions, 
which enforces the equivalence of the $x$ and $y$ directions.

\begin{table}[htb]
\caption{
The on-site energies (in eV) of the $d_{z^2}$, $d_{x^2-y^2}$, $d_{yz}$, $d_{xz}$,
and $d_{xy}$ Fe orbitals in the NM and Q$_M$ AFM phases in LaFeAsO and LaFePO.
In the Q$_M$ AFM phase, the on-site energies are shown separately for the spin up (majority spin) and spin down (minority spin) orbitals.
 }
\label{onsite-energy}
\begin{tabular}{|c|ccc|ccc|}
\hline
  & \multicolumn{3}{|c|}{LaFeAsO} & \multicolumn{3}{|c|}{LaFePO} \\
\hline
  & NM & \multicolumn{2}{c|}{Q$_M$} & NM & \multicolumn{2}{c|}{Q$_M$} \\
  &    & up    & dn    &    & up    & dn    \\
\hline
$z^2$     &  -0.11 & -0.95 & 0.18 & -0.17  & -0.35 & -0.04 \\
$x^2-y^2$ &  -0.27 & -1.14 & 0.07 & -0.27  & -0.44 & -0.14 \\
$yz$      &   0.02 & -0.67 & 0.23 & -0.04  & -0.19 &  0.07 \\
$xz$      &   0.02 & -0.70 & 0.21 & -0.04  & -0.21 &  0.07 \\
$xy$      &   0.18 & -0.50 & 0.40 &  0.23  &  0.13 &  0.30 \\
\hline
\end{tabular}
\end{table}

In the Q$_M$ AFM phase, the corresponding hopping parameters (both spin up and spin down) are 
either the same or
very close to the NM value, except for two cases. These differences are intimately related to
the changes in the corresponding WFs, as we now explain.
The first one is the $d_{xz}-d_{xz}$ hopping between parallel spin Fe neighbors ($x$ direction) of a majority spin
electron, whose absolute value increases significantly from the NM case (from -0.06 to -0.29 eV, 
see the highlighted numbers in Table \ref{LaFeAsO-hopping}). 
This opens an extra hopping channel in addition to
the original $d_{xz}-d_{xz}$ hopping in the $y$ direction.
In the NM state, the electrons in the $d_{xz}$ or $d_{yz}$ orbitals separately 
only hop in one direction (in the sense that the
hopping parameters in other directions are relatively small).
The dramatic change of the $3d_{xz}$ bands near Fermi level from NM to Q$_M$ AFM,
noted in several previous studies, can be 
traced to this difference.

The other case is for $d_{xy}-d_{xy}$ hopping, again between parallel spin atoms ($x$ direction). 
In the NM phase, the $d_{xy}-d_{xy}$ hoppings in $x$ and $y$ directions are the 
same by symmetry, with an amplitude of 0.18 eV. 
In the Q$_M$ AFM phase, 
this hopping in the $y$ direction for both spins is slightly enhanced to 0.23 eV. 
However, the $d_{xy}-d_{xy}$ hopping in the $x$ direction is significantly enhanced to 0.33 for the majority spin 
and suppressed to 0.07 for the minority spin. 
These differences shows that the broken symmetry
has a strong effect on the $d_{xy}$ orbital's environment.

\begin{table*}[htb]
\caption{The hopping parameters (in eV) 
of the Fe1 $3d_{yz}$, $3d_{xz}$, and $3d_{xz}$ orbitals 
to all the five $3d$ orbitals of its nearest neighbor Fe2 and Fe4 atoms and next-nearest neighbor Fe3 atom 
in the nonmagnetic and Q$_M$ AFM phases of LaFePO.
 }
\label{LaFePO-hopping}
\begin{tabular}{|cc|ccc|ccc|ccc|}
\hline
Fe1   &  & \multicolumn{3}{|c|}{$yz$} & \multicolumn{3}{|c|}{$xz$} & \multicolumn{3}{|c|}{$xy$} \\
\hline
      &  & NM & \multicolumn{2}{c|}{Q$_M$} & NM & \multicolumn{2}{c|}{Q$_M$} & NM & \multicolumn{2}{c|}{Q$_M$} \\
      &  &    & up    & dn    &    & up    & dn    &    & up    & dn \\
\hline
Fe2  & $z^2$     & -0.06 & -0.07 & -0.05 &  0    &  0    &  0  & 0 & 0 & 0  \\
     & $x^2-y^2$ &  0.42 &  0.44 &  0.41 &  0    &  0    &  0  & 0 & 0 & 0  \\
     & $yz$      & {\it -0.37} & {\it -0.37} & {\it -0.34} &  0    &  0    &  0  & 0 & 0 & 0   \\
     & $xz$      &  0    &  0    &  0    & -0.09 & {\bf -0.15} & -0.03 & -0.23 & -0.23 & -0.22 \\
     & $xy$      &  0    &  0    &  0    & -0.23 & -0.23 & -0.23 & {\bf -0.27} & {\bf -0.31} & {\bf -0.24}\\
\hline
Fe4  & $z^2$     &  0    &  0    &  0    & -0.06 & -0.06 & -0.06 & 0 & 0 & 0 \\
     & $x^2-y^2$ &  0    &  0    &  0    & -0.42 & -0.43 & -0.42 & 0 & 0 & 0 \\
     & $yz$      & -0.09 & -0.09 & -0.09 &  0    &  0    &  0   & -0.23 & -0.24 & -0.22 \\
     & $xz$      &  0    &  0    &  0    & {\it -0.36} & {\it -0.36} & {\it }-0.36 & 0 & 0 & 0 \\
     & $xy$      & -0.23 & -0.22 & -0.24 &  0    &  0    &  0   & -0.27 & -0.27 & -0.28 \\
\hline
Fe3  & $z^2$     & -0.09 & -0.08 & -0.08 & -0.09 & -0.09 & -0.08 & -0.24 & -0.24 & -0.24 \\
     & $x^2-y^2$ & -0.13 &  0.13 & -0.13 & -0.13 & -0.12 & -0.13  & 0 & 0 & 0 \\
     & $yz$      &  {\it 0.25} & {\it 0.25} & {\it 0.25} &  0.09 &  0.10 &  0.09 & 0.04 & 0.04 & 0.05 \\
     & $xz$      &  0.09 &  0.08 &  0.10 & {\it 0.25} & {\it 0.25} & {\it 0.25} & 0.04 & 0.04 & 0.05 \\
     & $xy$      & -0.04 & -0.05 & -0.04 & -0.04 & -0.05 & -0.04 & 0.16 & 0.16 & 0.16 \\
\hline
\end{tabular}
\end{table*}

The magnitude of the changes of the hopping parameters in the two special cases mentioned above,
and thus the magnetic order induced changes in WFs, 
is directly related to the magnitude of the ordered Fe magnetic moment in the Q$_M$ AFM state, which is evident
by comparing the case of LaFeAsO and LaFePO (see Table \ref{LaFeAsO-hopping} and \ref{LaFePO-hopping}).
The iron atom in the Q$_M$ AFM state in the former compound has a large ordered magnetic
moment of 1.9 $\mu_B$ while in the latter compound it is very weak, only 0.5 $\mu_B$, in DFT-LSDA calculations.
The difference in the ordered Fe magnetic moment is consistent with
the change of hopping parameters of $d_{xz}-d_{xz}$ and $d_{xy}-d_{xy}$ 
in the $x$ direction of the spin majority electron from the NM to the Q$_M$ AFM state, as shown
in Table \ref{LaFeAsO-hopping} and Table \ref{LaFePO-hopping}.

The difference in the changes of the hopping parameters of each Fe $3d$ orbital 
from NM phase to Q$_M$ AFM phase is related to the spin polarization 
of each orbital in the Q$_M$ AFM phase, 
as shown in Table \ref{spin-polarization}. 
The $3d_{xz}$ orbital has the largest moment (0.51 $\mu_B$ in LaFeAsO), followed by the
$3d_{xy}$ orbital (0.48 $\mu_B$ in LaFeAsO). 
The other three orbitals have significantly smaller moments (less than 0.41 $\mu_B$ in LaFeAsO). 
It is clear that the orbital with larger orbital spin magnetic moment has bigger changes in the relevant hopping parameters.
The difference of the relevant hopping parameters between LaFeAsO and LaFePO 
can also be traced to the difference in the orbital spin magnetic moment.

\begin{table*}[htb]
\caption{
Occupation numbers and spin polarizations in $3d$ 
orbitals in the NM and Q$_M$ AFM phases of LaFeAsO and LaFePO compounds. 
$\delta$n is the difference of the total occupation number in each orbital between the Q$_M$ AFM phase and the NM phase.
m is the spin magnetic moment in each orbital in the Q$_M$ AFM phase.
 }
\label{spin-polarization}
\begin{tabular}{|c|ccccc|ccccc|}
\hline
  & \multicolumn{5}{|c|}{LaFeAsO} & \multicolumn{5}{|c|}{LaFePO} \\
\hline
  & NM & \multicolumn{4}{c|}{Q$_M$} & NM & \multicolumn{4}{c|}{Q$_M$} \\
  &    & up    & dn  & $\delta$n   & m   & & up    & dn  & $\delta$n & m   \\
\hline
$z^2$     &  0.71 & 0.89 & 0.48 & -0.05 & 0.41 & 0.69 & 0.75 & 0.64 &  0.01 & 0.11 \\
$x^2-y^2$ &  0.57 & 0.80 & 0.45 &  0.10 & 0.34 & 0.54 & 0.59 & 0.50 &  0.01 & 0.09  \\
$yz$      &  0.65 & 0.85 & 0.57 &  0.11 & 0.28 & 0.67 & 0.72 & 0.64 &  0.01 & 0.08  \\
$xz$      &  0.65 & 0.86 & 0.35 & -0.10 & 0.51 & 0.67 & 0.75 & 0.56 & -0.03 & 0.19 \\
$xy$      &  0.68 & 0.88 & 0.39 & -0.11 & 0.48 & 0.67 & 0.71 & 0.62 & -0.00 & 0.09 \\
\hline
\end{tabular}
\end{table*}

The transition to the Q$_M$ AFM state is accompanied, in a local picture and to
second order, by an extra kinetic energy gain of
\begin{equation}
\Delta E_{xx} =-t_{xx}^2/(U'-J_H)
\end{equation} from the hopping process
of $d_{xz}-d_{xz}$ hopping in the $x$ direction, which is comparable with $\Delta E_{xy}$.
(Note that $\Delta E_{xx}$ is negligible in the NM state.) 
A substantial extra kinetic energy gain can also be obtained from the $d_{xy}-d_{xy}$ hopping in the $x$ direction.
 The anisotropy arises because the majority-spin electron in the $3d_{xz}$ orbital can
hop in both directions ({\it i.e.} to both parallel and antiparallel spin neighbors), while others 
in the $3d_{xz}$ and $3d_{yz}$ orbitals can basically only hop in one direction.
This anisotropy is reflected in a large symmetry lowering of the $3d_{xy}$ orbital in the AFM phase. 
The anisotropy leads to a large spin polarization (orbital spin magnetic moment) in the $3d_{xz}$ and $3d_{xy}$ orbital, 
which may also be related to the tetragonal to orthorhombic structural transition such that the lattice constant along the aligned-spin direction
($x$ direction in this paper) becomes shorter than the other direction ($y$ direction in this paper, thus $a<b$).

The additional $3d_{xz}-3d_{xz}$ hopping and the enhancement of the $3d_{xy}-3d_{xy}$ hoppings,  
both in the $x$ direction of the spin majority electron, promote kinetic energy gain.
However, as pictured in Fig. \ref{hopping}a, the $3d_{xz}$ spin up electron of Fe1 atom cannot hop
in the $x$ direction due to the Pauli principle. 
In order to take advantage of this extra kinetic energy gain of $\Delta E_{xx}$, 
the spin up occupation number of the $3d_{xz}$ orbital should not be unity but instead must fluctuate.
The same situation happens to the $3d_{xy}$ orbital.
The competition between the kinetic energy gain and Pauli principle 
results in a reduced magnetic moment and is possibly one mechanism of orbital fluctuation.

\section[Summary]{Summary}
In this paper, we have compared the electronic structures of LaFeAsO and LaFePO, in both NM and Q$_M$ AFM phases, 
and find that the stripe antiferromagnetism affects very differently the various Fe $3d$ orbital characters, 
even when the stripe antiferromagnetism is weak. 
By comparing LaFeAsO to LaFePO (and looking at similar results for other 1111 and 112 compounds\cite{Yin-thesis}), 
we find that the pnictide atom and the structure are influential in the 
formation of Q$_M$ AFM phase, consistent with several earlier reports that did not provide any
detailed analysis.
This information was obtained from a tight-binding representation for Fe $3d$ electrons 
based on first principles Wannier functions. 

In the nonmagnetic phase the electrons in Fe $3d_{xz}$ and $3d_{yz}$ orbitals have very different
amplitudes to hop in the $x$ and $y$ directions, resulting from the positions and chemical 
character of the pnictide atoms.  
Anti-intuitively, this ``anisotropy'' is almost gone for majority spin electrons in the AFM phase,  
when the $3d_{xz}$ (or $3d_{yz}$) electron can hop equally to parallel and antiparallel neighbors 
(both $x$ and $y$ directions).
This change is accompanied by a lowering of symmetry, and extension in space, in the $3d_{xy}$ 
Wannier function. 
The (large) changes in the near neighbor hopping parameters of the $3d_{xz}$ and $3d_{xy}$ orbitals in the $x$ direction  
is directly connected to the much larger orbital spin magnetic moments of these two orbitals than 
the other three orbitals. 

The anisotropy in hopping in the Fe $3d_{yz}$, $3d_{xz}$, and $3d_{xy}$ orbitals 
also favors orbital fluctuation by providing extra kinetic processes,
which are partly compensated by the Pauli principle which inhibits the hopping processes,
and which we expect to enhance fluctuations in the corresponding orbital occupation numbers (orbital fluctuation). 
Such fluctuations would reduce the ordered Fe magnetic moment in the Q$_M$ phase, bringing them closer to
the observed ordered moments. 
On the other hand, interlayer hoppings of the Fe $3d$ electrons
in the $z$ direction may help to stabilize the Fe magnetic moment in the Q$_M$ AFM phase.\cite{Yin-thesis}

\section[Acknowledgments]{Acknowledgments}
The authors thank Q. Yin and E. R. Ylvisaker for helpful discussions, 
and K. Koepernik for implementing the calculations of Wannier functions in FPLO code.
This work was supported by DOE grant DE-FG02-04ER46111.


\newpage

{\bf Supplementary Material to \\
Crystal Symmetry and Magnetic Order in Iron Pnictides: \\
a Tight Binding Wannier Function Analysis
}

Z. P. Yin and W. E. Pickett\\
Department of Physics, University of California Davis, Davis, CA 95616

\section[Calculational methods]{Calculational methods}
We perform first principle calculations using the full-potential local-orbital code\cite{FPLO} (FPLO8)
with local density approximation (LDA) exchange-correlation (XC) functional\cite{PW91} (PW92).
In the calculations, we use experimental lattice constants and internal atomic coordinates for the compounds
LaFeAsO, LaFePO, CaFeAsF, SrFeAsF, BaFe$_2$As$_2$, SrFe$_2$As$_2$, and CaFe$_2$As$_2$,
as used in our previous work\cite{Yin:arXiv0804.3355, Yin-njp, zpywep}.
For the other two hypothetical compounds LaFeNO and LaFeSbO, the lattice constants and internal atomic coordinates are taken from
the optimized equilibrium values of first principle calculations\cite{Yin-thesis}
done in the Q$_M$ AFM phase using GGA (PBE) XC functional\cite{PBE},
since such calculations were proven to predict the correct equilibrium lattice constants
and internal atomic coordinates compared to the experimental values in all the known iron pnictide compounds.\cite{zpywep, Yin-thesis}

\begin{table}[htb]
\caption{The ratios of hopping parameters t$_{xy}$, t$_{yx}$, t$_{xx}$ and t$_{yy}$ in the NM and
Q$_M$ AFM phases of a few iron-pnictides. }
\label{iron-hopping}
\begin{tabular}{|c|c|ccc|ccc|}
\hline
         &  & \multicolumn{3}{c}{$yz$} & \multicolumn{3}{|c|}{$xz$} \\
\hline
compound  &  & NM &\multicolumn{2}{c|}{Q$_M$} & NM & \multicolumn{2}{c|}{Q$_M$} \\
(mag. mom.)  &  &    & up  & dn  &   &  up  & dn \\
\hline
LaFeNO          & t$_{yx}$/t$_{xy}$ & -0.30 & -0.33 & -0.27 & -0.30 & -0.31 & -0.31 \\
(1.86 $\mu_B$)  & t$_{yy}$/t$_{xx}$ & -0.03 & -0.05 & -0.05 & -0.03 & -0.14 &  0.06 \\
\hline
LaFePO          & t$_{yx}$/t$_{xy}$ & -0.37 & -0.37 & -0.34 & -0.36 & -0.36 & -0.36 \\
(0.52 $\mu_B$)  & t$_{yy}$/t$_{xx}$ & -0.09 & -0.09 & -0.09 & -0.09 & -0.15 & -0.03 \\
\hline
LaFeAsO         & t$_{yx}$/t$_{xy}$ & -0.33 & -0.42 & -0.29 & -0.33 & -0.35 & -0.35 \\
(1.90 $\mu_B$)  & t$_{yy}$/t$_{xx}$ & -0.06 & -0.09 & -0.09 & -0.06 & -0.29 &  0.09 \\
\hline
LaFeSbO         & t$_{yx}$/t$_{xy}$ & -0.26 & -0.39 & -0.21 & -0.26 & -0.28 & -0.27 \\
(2.45 $\mu_B$)  & t$_{yy}$/t$_{xx}$ & -0.07 & -0.11 & -0.11 & -0.07 & -0.38 &  0.16 \\
\hline \hline
CaFeAsF         & t$_{yx}$/t$_{xy}$ & -0.36 & -0.43 & -0.34 & -0.36 & -0.37 & -0.37 \\
(1.75 $\mu_B$)  & t$_{yy}$/t$_{xx}$ & -0.06 & -0.08 & -0.08 & -0.06 & -0.27 &  0.08 \\
\hline
SrFeAsF         & t$_{yx}$/t$_{xy}$ & -0.35 & -0.43 & -0.31 & -0.35 & -0.37 & -0.37 \\
(1.96 $\mu_B$)  & t$_{yy}$/t$_{xx}$ & -0.08 & -0.10 & -0.10 & -0.08 & -0.31 &  0.08 \\
\hline \hline
BaFe$_2$As$_2$  & t$_{yx}$/t$_{xy}$ & -0.32 & -0.40 & -0.29 & -0.32 & -0.34 & -0.34 \\
(1.88 $\mu_B$)  & t$_{yy}$/t$_{xx}$ & -0.08 & -0.10 & -0.10 & -0.08 & -0.28 &  0.07 \\
\hline
SrFe$_2$As$_2$  & t$_{yx}$/t$_{xy}$ & -0.33 & -0.40 & -0.31 & -0.33 & -0.34 & -0.35 \\
(1.78 $\mu_B$)  & t$_{yy}$/t$_{xx}$ & -0.08 & -0.10 & -0.10 & -0.08 & -0.28 &  0.06 \\
\hline
CaFe$_2$As$_2$  & t$_{yx}$/t$_{xy}$ & -0.33 & -0.38 & -0.32 & -0.33 & -0.35 & -0.35 \\
(1.67 $\mu_B$)  & t$_{yy}$/t$_{xx}$ & -0.08 & -0.10 & -0.10 & -0.08 & -0.28 &  0.06 \\
\hline
\end{tabular}
\end{table}

\section[Further Observations]{Further Observations}
Similar hopping parameters compared to LaFeAsO have been obtained for the CaFeAsF, SrFeAsF,
and $M$Fe$_2$As$_2$ ($M$=Ba, Sr, Ca) compounds,
(which have similar FeAs layers), as shown in Table \ref{iron-hopping}.
However, replacing As in LaFeAsO with other pnictides (N, P and Sb) results in similar $t_{xy}$, $t_{yx}$ and $t_{yy}$ but different $t_{xx}$.
Compared to LaFeAsO, the $t_{xx}$ for the majority spin electron in the Q$_M$ AFM phase is reduced for LaFeNO and LaFePO,
but enhanced in LaFeSbO. The importance of the pnictide for the formation of the Q$_M$ AFM phase is evident.

Another important factor is the interlayer hoppings.
The interlayer distance of FeAs layers in 1111-compounds is in the range of 8.2 -9.0 $\AA$
and it is much smaller in 122-compounds, ranging from 5.9 $\AA$ to 6.5 $\AA$.
The interlayer hopping parameters of Fe $3d$ electrons in the $z$ direction are negligible in 1111-compounds
but become substantial for certain hoppings
in 122-compounds, especially in CaFe$_2$As$_2$, whose interlayer distance of FeAs layers is only 5.9 $\AA$.
Certain interlayer hopping parameters are as large as 0.15 eV for $3d_{xy}$ and $3d_{z^2}$ orbitals,
and 0.07 eV for $3d_{yz}$, $3d_{xz}$ and $3d_{x^2-y^2}$ orbitals,
calculated in the Q$_M$ AFM phase for CaFe$_2$As$_2$, which has the smallest interlayer distance.

The large interlayer hopping parameters for the Fe $3d_{xy}$ orbital,
which at first sight seems very surprising,
becomes clear by noting that the $3d_{xy}$ Wannier orbital
is strongly distorted from its symmetric atomic shape to its nearest neighbor As atoms above and below the Fe plane,
as shown in Fig.4 in the original paper.
This extension in the $z$ direction will favor interlayer hoppings,
especially when the interlayer distance is small, as in the case of CaFe$_2$As$_2$.
For comparison, the interlayer hopping parameters (if not zero) are less than 0.01 eV in LaFeAsO.
The increasing hopping of Fe $3d$ electrons in the $z$ direction increases the interlayer coupling,
and may inhibit fluctuations and thereby help to stabilize the ordered Fe magnetic moment in the Q$_M$ AFM phase.
The $k_z$ dispersion correlates with the experimental observations that the measured Fe magnetic moments in the Q$_M$ AFM phase
are significantly larger in 122-compounds ($\sim$ 0.9 $\mu_B$) than 1111-compounds ($\sim$ 0.4 $\mu_B$).

\end{document}